\documentclass[12pt]{article}
\newcommand{\nc}{\newcommand}
\nc{\renc}{\renewcommand}

\usepackage{calc}
\usepackage{ifthen}
\usepackage{graphicx}
\usepackage{epsfig}
\usepackage{miniplot}

\usepackage{wrapfig}
\usepackage{subfigure}
\usepackage{rotfloat}

\usepackage{epsfig}
\usepackage{graphicx}
\usepackage{epsf}

\nc{\half}{{\textstyle{1\over2}}}
\nc{\etal}{\mbox{\it et al. }}
\nc{\ie}{{\it i.e.}}
\nc{\eg}{{\it e.g.}}

\renc{\thefootnote}{\arabic{footnote}}
\nc{\capt}[1]{{\bf Figure.} {\small\sl #1}}

\nc{\eqs}[2]{\mbox{Eqs.~(\ref{#1},\,\ref{#2})}}
\nc{\eq}[1]{\mbox{Eq.~(\ref{#1})}}

\nc{\figs}[2]{\mbox{Figs.~(\ref{#1},\,\ref{#2})}}
\nc{\fig}[1]{\mbox{Fig~.(\ref{#1})}}

\nc{\tag}[1]{\label{#1} \marginpar{{\footnotesize #1}}}
\nc{\mtag}[1]{\label{#1} \mbox{\marginpar{{\footnotesize #1}}}}
\renc{\baselinestretch}{1.5}
\newlength{\overeqskip}
\newlength{\undereqskip}
\nc{\be}[1]{\begin{equation} \mbox{$\label{#1}$}}
\nc{\bea}[1]{\begin{eqnarray} \mbox{$\label{#1}$}}
\nc{\Section}[2]{\section{#2}\label{#1}}
\nc{\Bibitem}[1]{\bibitem{#1}}
\nc{\Label}[1]{\label{#1}}

\nc{\bep}[1]{\begin{equation} \mbox{$A.\label{#1}$}}

\nc{\eea}{\vspace{\undereqskip}\end{eqnarray}}
\nc{\ee}{\vspace{\undereqskip}\end{equation}}
\nc{\bdm}{\begin{displaymath}}
\nc{\edm}{\end{displaymath}}
\nc{\dpsty}{\displaystyle}
\nc{\bc}{\begin{center}}
\nc{\ec}{\end{center}}
\nc{\ba}{\begin{array}}
\nc{\ea}{\end{array}}
\nc{\bab}{\begin{abstract}}
\nc{\eab}{\end{abstract}}
\nc{\btab}{\begin{tabular}}
\nc{\etab}{\end{tabular}}
\nc{\bit}{\begin{itemize}}
\nc{\eit}{\end{itemize}}
\nc{\ben}{\begin{enumerate}}
\nc{\een}{\end{enumerate}}
\nc{\bfig}{\begin{figure}}
\nc{\efig}{\end{figure}}
\nc{\arreq}{&\!=\!&}
\nc{\arrmi}{&\!-\!&}
\nc{\arrpl}{&\!+\!&}
\nc{\arrap}{&\!\!\!\approx\!\!\!&}
\nc{\non}{\nonumber\\*}
\nc{\align}{\!\!\!\!\!\!\!\!&&}

\def\lsim{\; \raise0.3ex\hbox{$<$\kern-0.75em
      \raise-1.1ex\hbox{$\sim$}}\; }
\def\gsim{\; \raise0.3ex\hbox{$>$\kern-0.75em
      \raise-1.1ex\hbox{$\sim$}}\; }
\nc{\DOT}{\hspace{-0.08in}{\bf .}\hspace{0.1in}}
\nc{\Laada}{\hbox {$\sqcap$ \kern -1em $\sqcup$}}
\nc\loota{{\scriptstyle\sqcap\kern-0.55em\hbox{$\scriptstyle\sqcup$}}}
\nc\Loota{{\sqcap\kern-0.65em\hbox{$\sqcup$}}}
\nc\laada{\Loota}
\nc{\qed}{\hskip 3em \hbox{\BOX} \vskip 2ex}

\nc{\real}{{\rm I \! R}}
\nc{\Z}{{\sf Z \!\!\! Z}}
\nc{\complex}{{\rm C\!\!\! {\sf I}\,\,}}
\def\bigid{\leavevmode\hbox{\small1\kern-3.8pt\normalsize1}}
\def\id{\leavevmode\hbox{\small1\kern-3.3pt\normalsize1}}
\nc{\slask}{\!\!\!/}
\nc{\bis}{{\prime\prime}}
\nc{\pa}{\partial}
\nc{\na}{\nabla}
\nc{\ra}{\rangle}
\nc{\la}{\langle}
\nc{\goto}{\rightarrow}
\nc{\swap}{\leftrightarrow}

\nc{\EE}[1]{ \mbox{$\cdot10^{#1}$} }
\nc{\abs}[1]{\left|#1\right|}
\nc{\at}[2]{\left.#1\right|_{#2}}
\nc{\norm}[1]{\|#1\|}
\nc{\abscut}[2]{\Abs{#1}_{\scriptscriptstyle#2}}
\nc{\vek}[1]{{\rm\bf #1}}
\nc{\integral}[2]{\int\limits_{#1}^{#2}}
\nc{\inv}[1]{\frac{1}{#1}}
\nc{\dd}[2]{{{\partial #1}\over{\partial #2}}}
\nc{\ddd}[2]{{{{\partial}^2 #1}\over{\partial {#2}^2}}}
\nc{\dddd}[3]{{{{\partial}^2 #1}\over
        {\partial #2 \partial #3}}}
\nc{\dder}[2]{{{d #1}\over{d #2}}}
\nc{\ddder}[2]{{{d^2 #1}\over{d {#2}^2}}}
\nc{\dddder}[3]{{d^2 #1}\over
        {d #2 d #3}}
\nc{\dx}[1]{d\,^{#1}x}
\nc{\dy}[1]{d\,^{#1}y}
\nc{\dz}[1]{d\,^{#1}z}
\nc{\dl}[1]{\frac{d\,^{#1}l}{(2\pi)^{#1}}}
\nc{\dk}[1]{\frac{d\,^{#1}k}{(2\pi)^{#1}}}
\nc{\dq}[1]{\frac{d\,^{#1}q}{(2\pi)^{#1}}}

\nc{\cc}{\mbox{$c.c.$ }}
\nc{\hc}{\mbox{$h.c.$ }}
\nc{\cf}{cf.\ }
\nc{\erfc}{{\rm erfc}}
\nc{\Tr}{{\rm Tr\,}}
\nc{\tr}{{\rm tr\,}}
\nc{\pol}{{\rm pol}}
\nc{\sign}{{\rm sign}}
\nc{\bfT}{{\bf T }}

\def\GeV{{\rm\ GeV}}

\nc{\cA}{{\cal A}}
\nc{\cB}{{\cal B}}
\nc{\cD}{{\cal D}}
\nc{\cE}{{\cal E}}
\nc{\cG}{{\cal G}}
\nc{\cH}{{\cal H}}
\nc{\cL}{{\cal L}}
\nc{\cO}{{\cal O}}
\nc{\cT}{{\cal T}}
\nc{\cN}{{\cal N}}
\nc{\rvac}[1]{|{\cal O}#1\rangle}
\nc{\lvac}[1]{\langle{\cal O}#1|}
\nc{\rvacb}[1]{|{\cal O}_\beta #1\rangle}
\nc{\lvacb}[1]{\langle{\cal O}_\beta #1 |}
\nc{\bb}{\bar{\beta}}
\nc{\bt}{\tilde{\beta}}
\nc{\ctH}{\tilde{\cal H}}
\nc{\chH}{\hat{\cal H}}
\nc{\1}{\aa}
\nc{\2}{\"{a}}
\nc{\3}{\"{o}}
\nc{\4}{\AA}
\nc{\5}{\"{A}}
\nc{\6}{\"{O}}
\nc{\al}{\alpha}
\nc{\g}{\gamma}
\nc{\Del}{\Delta}
\nc{\e}{\epsilon}
\nc{\eps}{\epsilon}
\nc{\lam}{\lambda}
\nc{\om}{\omega}
\nc{\Om}{\Omega}
\nc{\ve}{\varepsilon}
\nc{\mn}{{\mu\nu}}
\nc{\kap}{\kappa}
\nc{\vp}{\varphi}

\nc{\advp}[3]{{\it  Adv.\ in\ Phys.\ }{{\bf #1} {(#2)} {#3}}}
\nc{\annp}[3]{{\it  Ann.\ Phys.\ (N.Y.)\ }{{\bf #1} {(#2)} {#3}}}
\nc{\apl}[3]{{\it  Appl. Phys. Lett. }{{\bf #1} {(#2)} {#3}}}
\nc{\apj}[3]{{\it  Ap.\ J.\ }{{\bf #1} {(#2)} {#3}}}
\nc{\apjl}[3]{{\it  Ap.\ J.\ Lett.\ }{{\bf #1} {(#2)} {#3}}}
\nc{\app}[3]{{\it Astropart.\ Phys.\ }{{\bf #1} {(#2)} {#3}}}
\nc{\cmp}[3]{{\it  Comm.\ Math.\ Phys.\ }{{ \bf #1} {(#2)} {#3}}}
\nc{\cqg}[3]{{\it  Class.\ Quant.\ Grav.\ }{{\bf #1} {(#2)} {#3}}}
\nc{\epl}[3]{{\it  Europhys.\ Lett.\ }{{\bf #1} {(#2)} {#3}}}
\nc{\ijmp}[3]{{\it Int.\ J.\ Mod.\ Phys.\ }{{\bf #1} {(#2)} {#3}}}
\nc{\ijtp}[3]{{\it Int.\ J.\ Theor.\ Phys.\ }{{\bf #1} {(#2)} {#3}}}
\nc{\jcap}[3]{{\it JCAP }{{\bf #1} {(#2)} {#3}}}
\nc{\jhep}[3]{{\it JHEP }{{\bf #1} {(#2)} {#3}}}
\nc{\jmp}[3]{{\it  J.\ Math.\ Phys.\ }{{ \bf #1} {(#2)} {#3}}}
\nc{\jpa}[3]{{\it  J.\ Phys.\ A\ }{{\bf #1} {(#2)} {#3}}}
\nc{\jpc}[3]{{\it  J.\ Phys.\ C\ }{{\bf #1} {(#2)} {#3}}}
\nc{\jap}[3]{{\it J.\ Appl.\ Phys.\ }{{\bf #1} {(#2)} {#3}}}
\nc{\jpsj}[3]{{\it J.\ Phys.\ Soc.\ Japan\ }{{\bf #1} {(#2)} {#3}}}
\nc{\lmp}[3]{{\it Lett.\ Math.\ Phys.\ }{{\bf #1} {(#2)} {#3}}}
\nc{\mpl}[3]{{\it  Mod.\ Phys.\ Lett.\ }{{\bf #1} {(#2)} {#3}}}
\nc{\ncim}[3]{{\it  Nuov.\ Cim.\ }{{\bf #1} {(#2)} {#3}}}
\nc{\np}[3]{{\it  Nucl.\ Phys.\ }{{\bf #1} {(#2)} {#3}}}
\nc{\npps}[3]{{\it  Nucl.\ Phys.\ Proc.\ Suppl.\ }{{\bf #1} {(#2)} {#3}}}
\nc{\pr}[3]{{\it Phys.\ Rev.\ }{{\bf #1} {(#2)} {#3}}}
\nc{\pra}[3]{{\it  Phys.\ Rev.\ A\ }{{\bf #1} {(#2)} {#3}}}
\nc{\prb}[3]{{\it  Phys.\ Rev.\ B\ }{{{\bf #1} {(#2)} {#3}}}}
\nc{\prc}[3]{{\it  Phys.\ Rev.\ C\ }{{\bf #1} {(#2)} {#3}}}
\nc{\prd}[3]{{\it  Phys.\ Rev.\ D\ }{{\bf #1} {(#2)} {#3}}}
\nc{\prl}[3]{{\it Phys.\ Rev.\ Lett.\ }{{\bf #1} {(#2)} {#3}}}
\nc{\pl}[3]{{\it  Phys.\ Lett.\ }{{\bf #1} {(#2)} {#3}}}
\nc{\prep}[3]{{\it Phys.\ Rep.\ }{{\bf #1} {(#2)} {#3}}}
\nc{\prsl}[3]{{\it Proc.\ R.\ Soc.\ London\ }{{\bf #1} {(#2)} {#3}}}
\nc{\ptp}[3]{{\it  Prog.\ Theor.\ Phys.\ }{{\bf #1} {(#2)} {#3}}}
\nc{\ptps}[3]{{\it  Prog\ Theor.\ Phys.\ suppl.\ }{{\bf #1} {(#2)} {#3}}}
\nc{\physa}[3]{{\it  Physica\ A\ }{{\bf #1} {(#2)} {#3}}}
\nc{\physb}[3]{{\it  Physica\ B\ }{{\bf #1} {(#2)} {#3}}}
\nc{\phys}[3]{{\it Physica\ }{{\bf #1} {(#2)} {#3}}}
\nc{\rmp}[3]{{\it  Rev.\ Mod.\ Phys.\ }{{\bf #1} {(#2)} {#3}}}
\nc{\rpp}[3]{{\it Rep.\ Prog.\ Phys.\ }{{\bf #1} {(#2)} {#3}}}
\nc{\sjnp}[3]{{\it Sov.\ J.\ Nucl.\ Phys.\ }{{\bf #1} {(#2)} {#3}}}
\nc{\spjetp}[3]{{\it Sov.\ Phys.\ JETP\ }{{\bf #1} {(#2)} {#3}}}
\nc{\yf}[3]{{\it Yad.\ Fiz.\ }{{\bf #1} {(#2)} {#3}}}
\nc{\zetp}[3]{{\it Zh.\ Eksp.\ Teor.\ Fiz.\  }{{\bf #1}  {(#2)} {#3}}}
\nc{\zp}[3]{{\it Z.\ Phys.\ }{{\bf #1} {(#2)} {#3}}}
\nc{\ibid}[3]{{\sl ibid.\ }{{\bf #1} {#2} {#3}}}
\nc{\rf}[1]{(\ref{#1})}
\nc{\nn}{\nonumber \\*}
\nc{\bfB}{\bf{B}}
\nc{\bfv}{\bf{v}}
\nc{\bfx}{\bf{x}}
\nc{\bfy}{\bf{y}}
\nc{\vx}{\vec{x}}
\nc{\vy}{\vec{y}}
\nc{\oB}{\overline{B}}
\nc{\oI}{\overline{I}}
\nc{\oR}{\overline{R}}
\nc{\rar}{\rightarrow}
\nc{\ti}{\times}
\nc{\slsh}{\hskip-5pt/}
\nc{\sm}{Standard~Model~}
\nc{\MP}{M_{\rm Pl}}
\nc{\tp}{t_{\rm Pl}}
\nc{\ave}{\bar{E}}

\nc{\eff}{{\rm eff}}
\nc{\kk}{\vek{k}}
\nc{\pp}{{\rm p}}
\nc{\ga}{g_{a\gamma}}
\nc{\vv}{\\}
\nc{\eee}{{\bf E}}
\nc{\bbb}{{\bf B}}
\nc{\qcd}{T_{\rm QCD}}
\nc{\G}{\rm \ G}
\def\vec#1{{\bf #1}}

 \def\gae{\; ^{>}_{\sim} \;} 

\def\ell{e^{c}LL}

\begin{document}

{\title{\vskip-2truecm{\hfill {{\small \\
	\hfill \\
	}}\vskip 1truecm}
{\LARGE Simulations of the End of Supersymmetric Hybrid Inflation
 and Non-Topological Soliton Formation}}
\vspace{1cm}
{\author{
{\sc  Matt Broadhead$^{1}$ and John McDonald$^{2}$ \vspace{0.5cm}}\\
{\sl\small Theoretical Physics Division,}\\
{\sl\small Dept. of Mathematical Sciences, University of Liverpool,
Liverpool L69 3BX, England $^{1}$}
\vspace{0.5cm} 
\\ {\sl\small Dept. of Physics, University of Lancaster,
Lancaster LA1 4YB, England $^{2}$} \vspace{0.5cm}
}

\maketitle
\begin{abstract}
\noindent
        
          We present two- and three-dimensional
simulations of the growth of quantum
fluctuations of the scalar fields in supersymmetric hybrid inflation
models. For a natural range of couplings, sub-horizon quantum fluctuations undergo rapid
growth due to scalar field dynamics, resulting in the
formation of quasi-stable non-topological solitons (inflaton condensate lumps)
which dominate the post-inflation era.

\end{abstract}
\vfil
\footnoterule
{\small $^1$mattb@amtp.liv.ac.uk}
{\small $^2$j.mcdonald@lancaster.ac.uk}

\thispagestyle{empty}
\newpage
\setcounter{page}{1}

\section{Introduction}

          Hybrid inflation models \cite{hi} are a promising class of model,   
having a classically flat inflaton potential without requiring fine-tuning of
coupling constants. 
In order to control radiative corrections to the inflaton potential, supersymmetric (SUSY) 
hybrid inflation models are favoured \cite{dti,fti,lr}. D-term hybrid inflation
models are of particular interest, since they naturally account for the absence of order $H^2$
corrections to the inflaton mass squared term coming from non-renormalisable 
interactions with non-zero F-terms. 
However, in spite of requiring fine-tuning to eliminate order $H^2$ corrections, 
several F-term hybrid inflation models have also been 
proposed. In light of recent reconsideration of the issue of fine-tuning 
in SUSY models and the idea of 'split supersymmetry' \cite{ss}, such models may 
still play an important role. 

      An important issue is the process by which inflation ends and the 
Universe reheats. It is known that the end of inflation and reheating in hybrid inflation 
is dominated by the non-linear dynamics of the scalar fields responsible for inflation 
\cite{tp1,tp2,icf}. Sub-horizon 
quantum fluctuations are rapidly amplified by scalar field dynamics once the symmetry
breaking transition ending inflation occurs, becoming effectively classical modes \cite{tp1,tp2}.
The energy density at the end of inflation 
rapidly becomes dominated by spatial fluctuations of the inflaton.  
For the case of D-term hybrid inflation, the rapid growth of the perturbations
is expected to dominate the energy density for $\lambda \gae 0.1g$ 
\cite{mb1}.  Similar dynamics may be observed in models of tachyonic inflation \cite{tachy}.
\footnote{For related analyses of tachyonic preheating see \cite{add}.}

       What is the subsequent evolution of this inhomogeneous
scalar field? There are two possible outcomes. The growth 
of the spatial perturbations of the inflaton 
could result in scalar field waves which scatter from each
other, dispersing the energy of the initially highly inhomogeneous state \cite{tp1,tp2}. 
Alternatively, the growth of the spatial perturbations might continue, resulting
in the formation of non-topological solitons corresponding to
inflaton breathers \cite{icf} (also known as oscillons \cite{osc}). 
This was first suggested in an alternative view of the end of
 hybrid inflation, inflaton condensate fragmentation \cite{icf}, which 
was based upon perturbing a homogeneous inflaton condensate assumed
 to form immediately after the end of hybrid inflation.  In \cite{icf} the 
non-topological solitons were called
inflaton condensate lumps. Their existence can be understood as the result of 
an effective attractive interaction between the inflaton scalars due to 
the symmetry breaking field \cite{icf}. 
Although the assumption of a homogeneous inflaton condensate is violated
by the rapid growth of quantum fluctuations within a few coherent oscillation periods,
the possibility that spatial perturbations could form non-topological solitons remains.
Such objects were subsequently observed in a
numerical simulation of the phase transition in hybrid inflation \cite{cpr}.
Oscillons such as inflaton condensate lumps are known to be quasi-stable,
having a lifetime much longer than the 
period of the inflaton coherent oscillations \cite{osc}. They decay by a slow 
radiation of scalar field waves. For the case of the effective theory associated with 
SUSY hybrid inflation models, the lifetime of the inflaton condensate lumps 
is $10^{3}-10^{4}$ times the oscillation period \cite{iball}.

              The possibility of non-topological soliton formation at the end of 
SUSY hybrid inflation is strengthened by the existence of stable Q-ball solutions of the
scalar field equations of D- and F-term inflation \cite{qballs}.
These are two-field Q-ball solutions, composed of a complex inflaton
field and a real symmetry breaking field. They carry a 
conserved global charge associated with the inflaton.  As a result, it is possible that
SUSY hybrid inflation could end by initially forming neutral non-topological solitons 
which subsequently decay to pairs of stable Q-balls of opposite global charge.
The formation of neutral condensate lumps
which decay to Q-ball pairs has been observed numerically 
in the context of Affleck-Dine condensates along MSSM flat directions \cite{kk}. 
(See also \cite{enq}.) It occurs because the initially neutral condensate lumps
are unstable with respect to growth of perturbations of the phase of the complex
scalar field, resulting in the formation of an energetically preferred state consisting of 
a stable Q-ball, anti-Q-ball pair. This is likely to
be a generic feature of the evolution of quasi-stable neutral
condensate lumps whenever a related stable Q-ball solution exists.

       Thus there exists the possibility that the post-inflation era of SUSY hybrid inflation 
models will be dominated by Q-balls containing most of the energy of the Universe.
Whether this happens depends upon whether the inhomogeneous state of the Universe 
immediately following inflation evolves into quasi-stable neutral condensate lumps or
simply disperses as scalar field waves. In this paper we report the results of   
a numerical simulation of the growth and evolution of quantum 
fluctuations of the scalar fields in SUSY hybrid inflation models. 
The growth of quantum fluctuations into classical fluctuations is studied using the
equivalent classical stochastic field method \cite{ps}, which replaces
the quantum field theory by an equivalent classical field theory with a classical 
probability distribution for the initial conditions. We focus on the case of D-term 
inflation. The results for F-term inflation are likely to be similar to those of
D-term inflation for a particular choice of the D-term inflation model couplings.

      The paper is organised as follows. In Section 2 we review SUSY hybrid inflation models
and the scalar field equations. In Section 3 we discuss the initial conditions and details of the 
numerical simulation. In Section 4 we present the results of 2-D and 3-D simulations of the
evolution of the scalar fields. 
In Section 5 we discuss a scaling property of the D-term inflation field equations
and the dependence of our numerical results on the coupling constants. 
In Section 6 we compare our results with earlier numerical simulations. 
In Section 7 we present our conclusions. 
In the Appendix we review the equivalent classical stochastic field method.

\section{SUSY Hybrid Inflation Models}

     The minimal D-term inflation model is described by the superpotential \cite{dti}
\be{w1} W = \lambda S \Phi_{+} \Phi_{-}    ~,\ee
where $S$ is the inflaton, $\Phi_{\pm}$ are fields with charges $\pm 1$ with respect to a Fayet-Iliopoulos 
$U(1)_{FI}$ gauge symmetry, $\xi > 0$ is the Fayet-Iliopoulos term ($\xi^{1/2} = 8.5 \times 10^{15} \GeV$ \cite{lr}) 
and $g$ is the $U(1)_{FI}$ gauge coupling. 
The scalar potential is given by 
\be{w2} V = \lambda^{2}|S|^{2}\left(|\Phi_{+}|^{2} + |\Phi_{-}|^{2}\right) 
+ \lambda^{2} |\Phi_{+}|^{2} |\Phi_{-}|^{2} + \frac{g^{2}}{2} \left(|\Phi_{+}|^{2}
- |\Phi_{-}|^{2} + \xi\right)^{2}    ~.\ee
Once $|S| < s_{c}/\sqrt{2} = g \xi^{1/2}/\lambda$, $\Phi_{-}$ develops a non-zero expectation value,
breaking the $U(1)_{FI}$ symmetry and ending inflation.      

    The simplest F-term inflation model is described by the superpotential \cite{fti}
\be{w3} W = \frac{\eta}{2}S \left(\Phi^{2} - \mu^{2}\right)       ~,\ee
where $S$ is the inflaton and $\Phi$ is the field 
which terminates inflation. $\mu^{2}$ and $\eta$ may be chosen 
real and positive. The scalar potential is then 
\be{w4} V = \eta^{2} |S|^{2} |\Phi|^{2} + \frac{\eta^{2}}{4} |\Phi^{2} - \mu^{2}|^{2}
~.\ee
For the case of real scalar fields, the F-term scalar field equations become equivalent to the
D-term equations in the limit where $\Phi_{+} = 0$ and $\lambda = \sqrt{2} g$.

    The superpotentials \eq{w1} and \eq{w3} are invariant with respect to an
R-symmetry under which only $S$ transforms, 
which manifests itself as a global $U(1)$ symmetry of the scalar potential, $S \rightarrow e^{i \alpha}S$. 
It is this symmetry which is responsible for the conserved global charge of the 
hybrid inflation Q-balls \cite{qballs}.

       In the following we will focus on the minimal D-term inflation model. We expect the end of inflation 
in F-term inflation models to be similar to D-term inflation in the case $\lambda = \sqrt{2}g$,
up to the effect of $U(1)_{FI}$ symmetry breaking and cosmic string formation, which our results indicate are
of secondary importance to the evolution of the energy density.  
During inflation, the inflaton evolves due to the 1-loop effective potential, 
$\Delta V(|S|,|\Phi_{-}|)$ \footnote{
$\Delta V(|S|,|\Phi_{-}|)$, has been calculated in \cite{mb1}, where $\Delta V$ 
is given for $s \equiv \sqrt{2} Re(S)$ and $\phi_{-} \equiv \sqrt{2} Re{\Phi_{-}}$. The complex version
is given by replacing $s$, $\phi_{-}$ by $\sqrt{2}|S|$, $\sqrt{2}|\Phi_{-}|$ in $\Delta V(s,\phi_{-})$.}.    
This determines its rate of rolling at the end of inflation, an important initial condition 
for the subsequent non-linear growth of the field perturbations. 
We assume throughout that $\Phi_{+} = 0$. This is justified since the 
$\Phi_{+}$ field has a large positive mass squared both during and after inflation. 
The equations of motion are then 
\be{e7} \ddot{S} + 3 H \dot{S} - \frac{\vec{\nabla}^{2}}{a^{2}}S    
= -\lambda^{2} |\Phi_{-}|^{2}S
- \frac{\partial \Delta V}{\partial S^{\dagger}}      ~,\ee
and
\be{e8} \ddot{\Phi}_{-} + 3 H \dot{\Phi}_{-} - \frac{\vec{\nabla}^{2}}{a^{2}}\Phi_{-}    
= -\lambda^{2} |S|^{2}\Phi_{-} + g^{2}\left(\xi - |\Phi_{-}|^{2} \right)
\Phi_{-} - \frac{\partial \Delta V}{\partial \Phi_{-}^{\dagger}}      ~.\ee

\section{Initial Conditions and Numerical Methods}

          The mechanism behind the formation of spatial inhomogeneities of the 
inflaton field at the end of D-term hybrid inflation is the 
dynamical amplification of sub-horizon quantum fluctuations of the scalar fields. In
the Appendix we review the formation of classical scalar field perturbations from  
dynamical amplification of quantum fluctuations. 
The process can be summarised as follows. At the end of hybrid inflation, the symmetry breaking
transition results in the formation of an effectively tachyonic scalar potential for the inflaton. 
Rapid growth of the quantum modes in this potential results in semi-classical quantum fluctuations. 
If the quantum fluctuations grow to become semi-classical whilst the field equations are
linear in the inflaton, then the quantum field theory becomes equivalent to solving the 
classical field equations with a classical probability distribution for the initial conditions, 
given by the Wigner function of the field modes
and their conjugate momenta in the semi-classical limit. The subsequent 
evolution of the scalar field fluctuations can then be studied by 
solving the {\it classical} scalar field equations with stochastic initial conditions 
(equivalent classical stochastic field).  In this way
 we can simulate the growth of sub-horizon quantum fluctuations
 at the end of hybrid inflation.

\subsection{Initial Conditions}

      The initial conditions for the simulation are obtained from
the values of $\phi(t)$ and $\dot{\phi}(t)$
for the homogeneous zero modes and values of the equivalent  classical
momentum modes $q(\vec{k},t)$
and $p(\vec{k},t)$. (Here $\phi$ represents any real scalar field.)  
The classical initial conditions are applied at the end of hybrid inflation, 
$|S|  =  s_{c}/\sqrt{2}$, at which time the $S$ and $\Phi_{-}$ fields are massless.

The fields are studied in a comoving box of volume $V = L^3$ with periodic boundary conditions, 
with the classical field $\phi$ and its derivative, $\dot{\phi} $, being expanded in terms of modes as  
\be{fe3}  \phi(\vec{x}, t) = \frac{1}{\sqrt{V}} \sum q(\vec{k},t) e^{i \vec{k}.\vec{x}}   ~\ee
and
\be{fe4}  \dot{\phi}(\vec{x}, t) = \frac{1}{\sqrt{V}} \sum p(\vec{k},t) e^{i \vec{k}.\vec{x}}   ~,\ee
where $\vec{k} = \Delta k (\alpha,\beta,\gamma)$ with integer $\alpha$, $\beta$, $\gamma$  
and $\Delta k = 2 \pi/L$.  

    For a massless scalar field in de Sitter space, the initial condition for non-zero modes $q(\vec{k},t)$
at $t=0$ corresponds to a complex Gaussian distribution for $|q(\vec{k},0)|$ with root mean
squared (r.m.s.) value given by (Appendix) \cite{ps}   
\be{fe5} |q(\vec{k}, 0)|_{rms} = \frac{1}{\sqrt{2 k}}\left(1 + \frac{1}{k^{2} \eta^{2}}\right)^{1/2}     ~,\ee
and a random phase factor $e^{i \delta_{\vec{k}}}$ for each $\vec{k}$. 
Here $\eta$ is the conformal time at the end of inflation ($\eta = -H^{-1}$) and $k = |\vec{k}|$. 
Since $\phi(\vec{x}, t)$ is a real scalar field, the modes and phase satisfy 
$q(-\vec{k}, t) = q^{\dagger}(\vec{k}, t)$
and $\delta_{-\vec{k}} = - \delta_{\vec{k}}$. The 
initial value of the mode function $p(\vec{k},t)$ has an r.m.s. magnitude 
\be{fe6} |p(\vec{k}, 0)|_{rms} = \frac{1}{\sqrt{2 k}|\eta|}\left(1 + \frac{1}{k^{2} 
\eta^{2}}\right)^{-1/2}       ~.\ee
For the sub-horizon modes of interest, $|k\eta| \equiv |k/H| \gg 1$.
In this case, from \eq{fe5} and \eq{fe6}, 
initially $|p(\vec{k}, 0)| \equiv |\dot{q}(\vec{k},0)|  \approx H |q(\vec{k},0)|$. 
Since the mode $q(\vec{k},t)$, corresponding to a wavefunction of wavenumber $\vec{k}$, 
will satisfy $\dot{q}(\vec{k},t) \approx k q(\vec{k},t) \gg H |q(\vec{k},0)|$
 at subsequent times, 
the initial value of $\dot{q}(\vec{k},t)$ from \eq{fe6} will have a negligible effect. 

        In our simulations we modelled the initial mode function $q(\vec{k},0)$ by
the r.m.s. value of $|q(\vec{k},0)|$ multiplied by a random complex phase for each $\vec{k}$.
To simulate the evolution of the 
scalar field we considered a spatial lattice with $\vec{x} = \Delta x (i,j,k)$ for 
integer $i,j,k$.  The sum over non-zero modes in \eq{fe3} then gives the 
initial spatial perturbation on the lattice,  
\be{fe7}  \delta \phi(\vec{x}, 0)  = 
\frac{1}{\sqrt{V}} \sum \frac{8}{\sqrt{2 k_{\alpha \beta \gamma}}} 
\cos\left( \Delta k \Delta x \alpha i + \theta_{\alpha} \right) 
\cos\left( \Delta k \Delta x \beta j + \theta_{\beta} \right)
\cos\left( \Delta k \Delta x \gamma k + \theta_{\gamma} \right)   ~,\ee
where $k_{\alpha\beta\gamma} = \Delta k(\alpha^{2}+\beta^{2}+\gamma^{2})^{1/2}$ and 
we have substituted $\delta_{\vec{k}} \rightarrow \theta_{\alpha}
 + \theta_{\beta} + \theta_{\gamma}$, 
where $\theta_{\alpha,\beta,\gamma}$ are random phases. 
We then solved the scalar field equations numerically on the lattice using
\eq{fe7} and $\delta \dot{\phi}(\vec{x},0) = 0$ for each of the real scalar fields.

     Only those modes which are amplified,
such that their occupation number becomes large, can be considered classical.  
Modes which are not amplified remain 
as quantum fluctuations. In simulations the quantum fluctuations
must be regularised, since summing over large $k$ modes will result in 
a large initial amplitude for the spatial perturbations, 
causing the assumption of linearity of the field equations 
to break down. We introduce a cut-off $\Lambda$ at a value  
given by the mass of the inflaton field in vacuum, $\Lambda = m_{S} = \lambda \xi^{1/2}$. 
This should be close to the largest value of $k$ 
for which modes can grow dynamically, $k_{max}$, although the exact
value of $k_{max}$ cannot be known a priori since it is determined by the full scalar field dynamics.

\section{Results of Numerical Simulations}

     In order to investigate the formation of neutral inflaton
condensate lumps, we have studied a model with a real inflaton field, $s \equiv \sqrt{2}Re(S)$. 
We considered numerical simulations in two- and three-dimensions, 
using a staggered leapfrog routine and the initial conditions for quantum
fluctuations  \eq{fe7}  applied to $s$, $\phi_{-\;1}$ and $\phi_{-\;2}$ at $s = s_{c}$, 
where $\Phi_{-} = (\phi_{-\;1} + i \phi_{-\;2})/\sqrt{2}$.

      In Figure 1, we show the results for the growth of the energy density in 
a 2-D simulation using a $200 \times 200$ lattice, for the case $\lambda = 0.14$ and $g=1$, for four
different times. We compare the energy density $\rho$ with the mean energy density averaged
 over the lattice, $\overline{\rho}$. 
The time from the first to last figure 
corresponds to five $S$ coherent oscillation periods, $\tau_{S}$,
where $\tau_{S} = 2 \pi m_{S}^{-1}$. 
The growth of the energy density perturbations to
non-linearity occurs within the first two coherent oscillation periods. We find that roughly circular 
lumps of energy density form, which are stable over the duration of the simulation.   
Essentially all of the energy density becomes concentrated in the 
lumps. We also observe some condensate lumps coalescing into 
larger lumps, in particular by comparing Figures 1c and 1d. 
The lattice spatial dimension is given by $L = 20 \pi m_{S}^{-1}$, which is 
small compared with the horizon $H^{-1} \approx (3/4 \pi)^{1/2} 
\lambda^{2} M_{Pl}/(g m_{S}^{2})$.

\begin{arrangedFigure}{1}{2}{arrangedO521}
{\footnotesize{
2D growth of energy density perturbations for $\lambda = 0.14$ and $g = 1$
on a 200 $\times$ 200 lattice. Black areas correspond to $\rho > 4 \overline{\rho}$.   
}}
   \subFig[$t = \tau_{S}$]{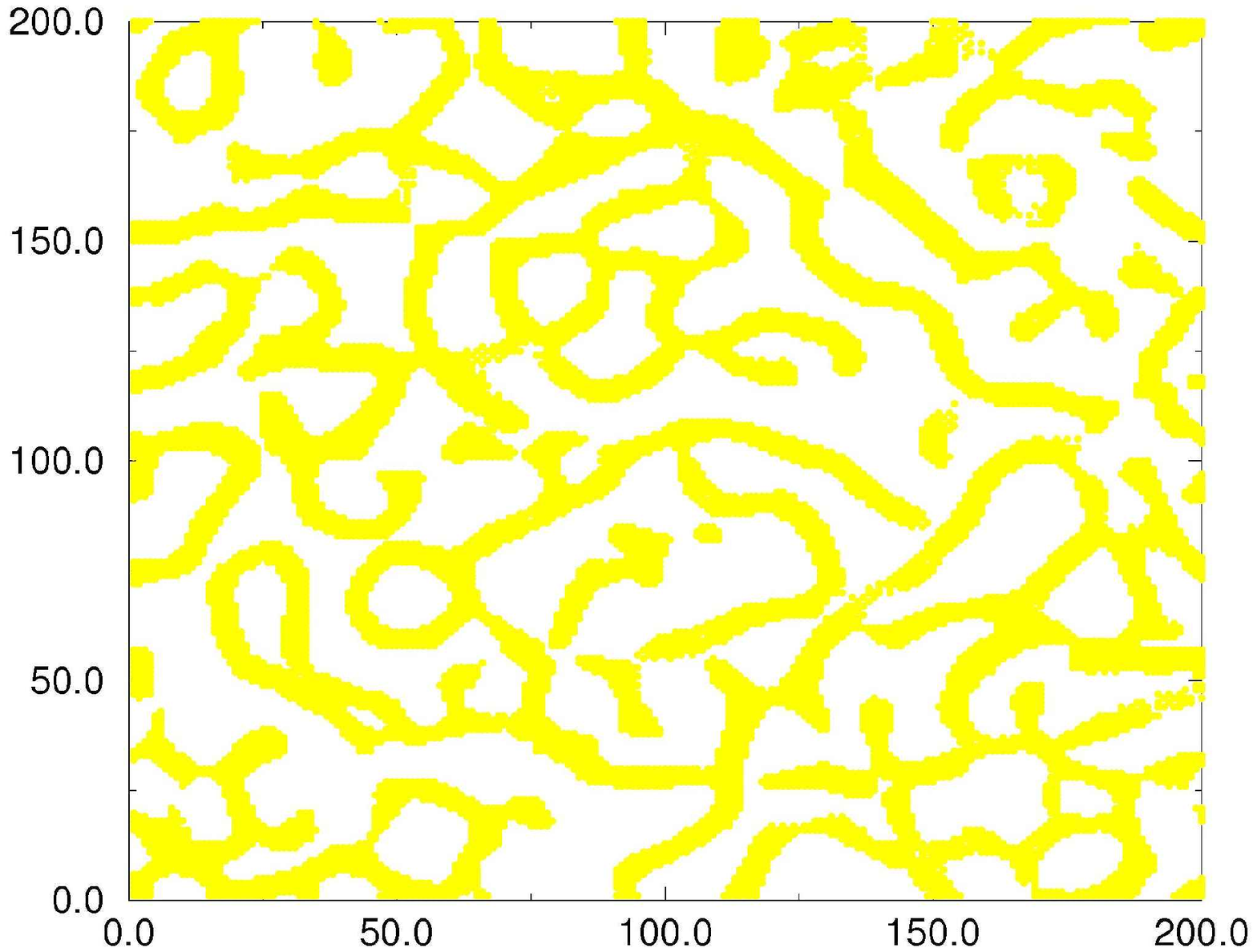}%
    \subFig[$t= 3 \tau_{S}$]{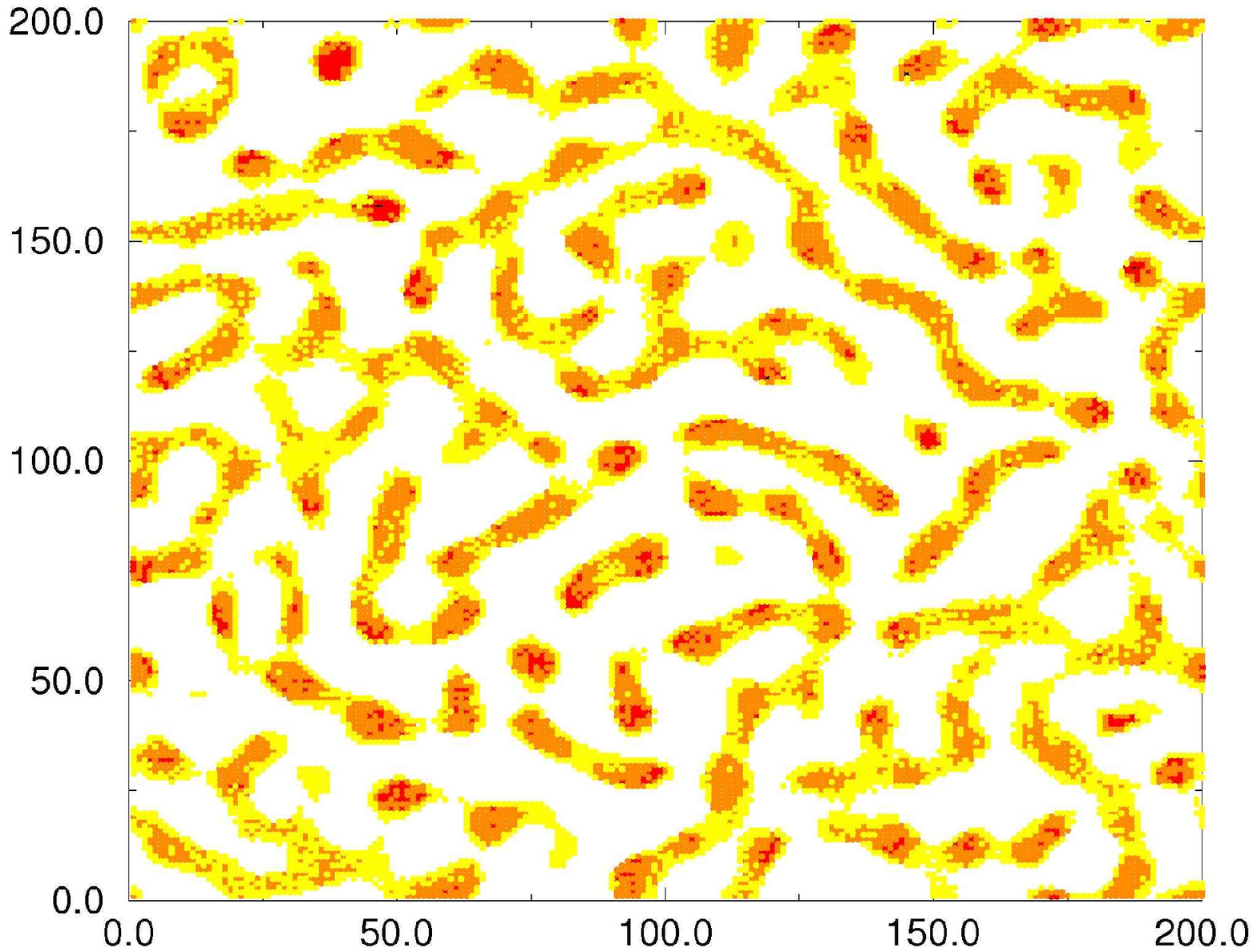}%
    \subFig[$t = 4 \tau_{S}$]{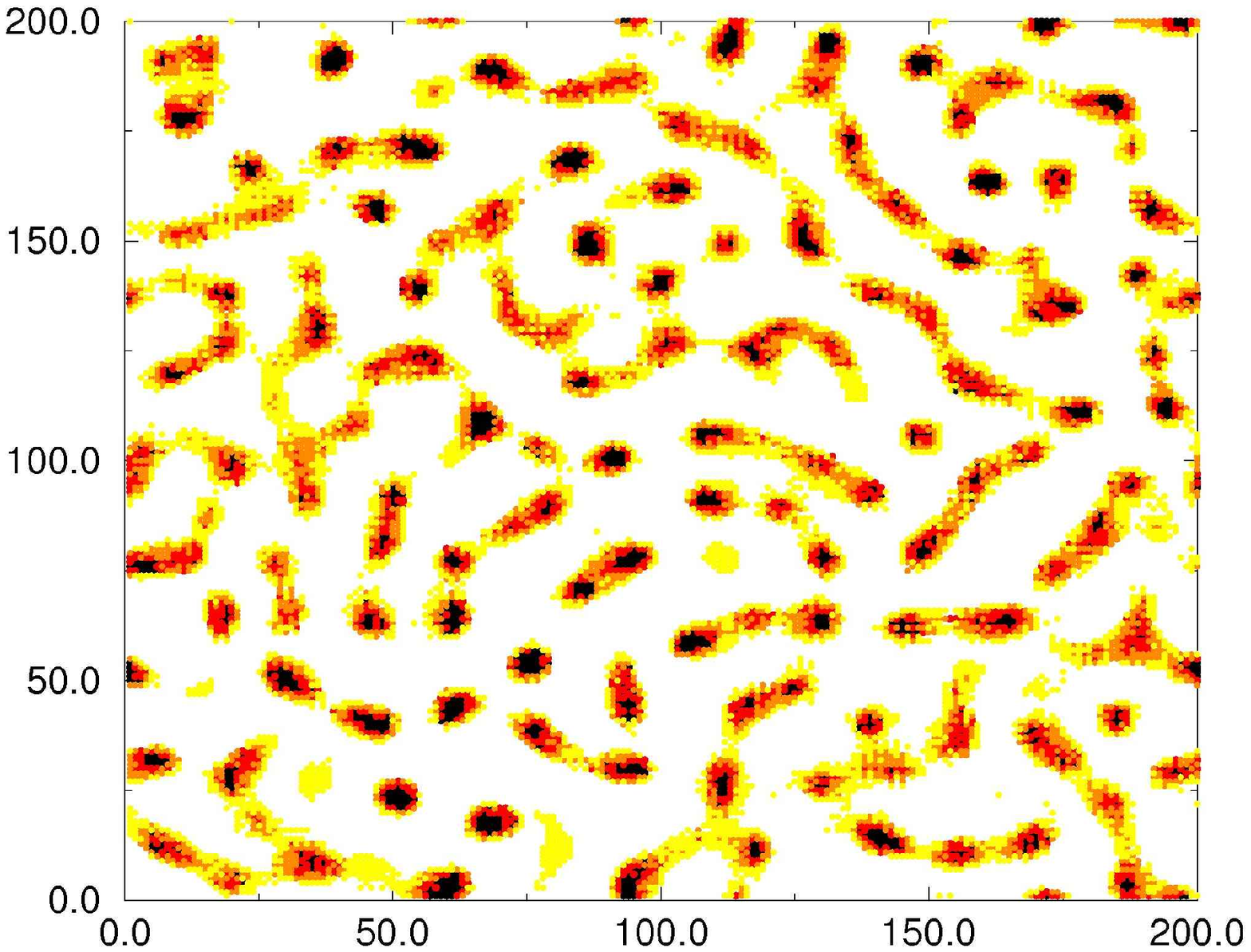}%
    \subFig[$t = 5 \tau_{S}$]{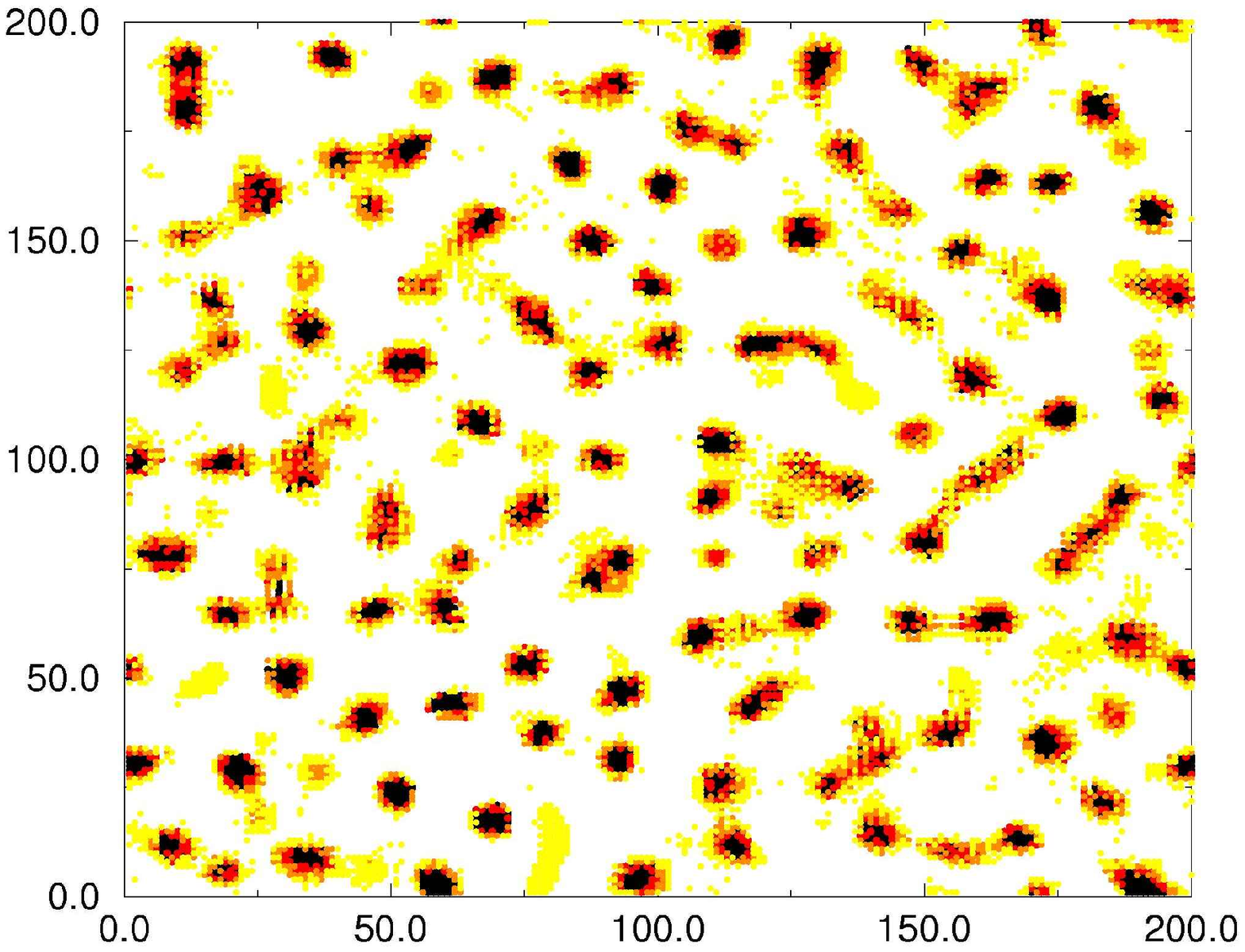}%
\setlength{\subFigureAboveCaptionSpace}{7mm} 
\end{arrangedFigure} 
\setlength{\subFigureAboveCaptionSpace}{0mm}

              In Figure 2, we show the amplitude of the $S$ field at the
centre of one of the energy density lumps in Figure 1. The oscillations of an 
unperturbed homogeneous inflaton field are shown for comparison. 
The field inside the lump is oscillating
in time, showing that the inflaton condensate lump is indeed a quasi-stable oscillon. 
This is important, as it is not obvious that the energy density lumps in Figure 1
are not simply concentrations
of the energy density with no coherent structure. If this were true, it would be possible for 
such concentrations to have a short lifetime due to a large amount of
kinetic energy associated with their constituent scalar particles. The fact the energy density lumps are 
oscillons suggests that they will have a very long 
lifetime compared with the coherent oscillation period, 
as observed in numerical and analytical studies of single-field oscillons \cite{osc,iball}.

\begin{figure}[htbp]
\begin{center}
\includegraphics[width=0.75\textwidth]{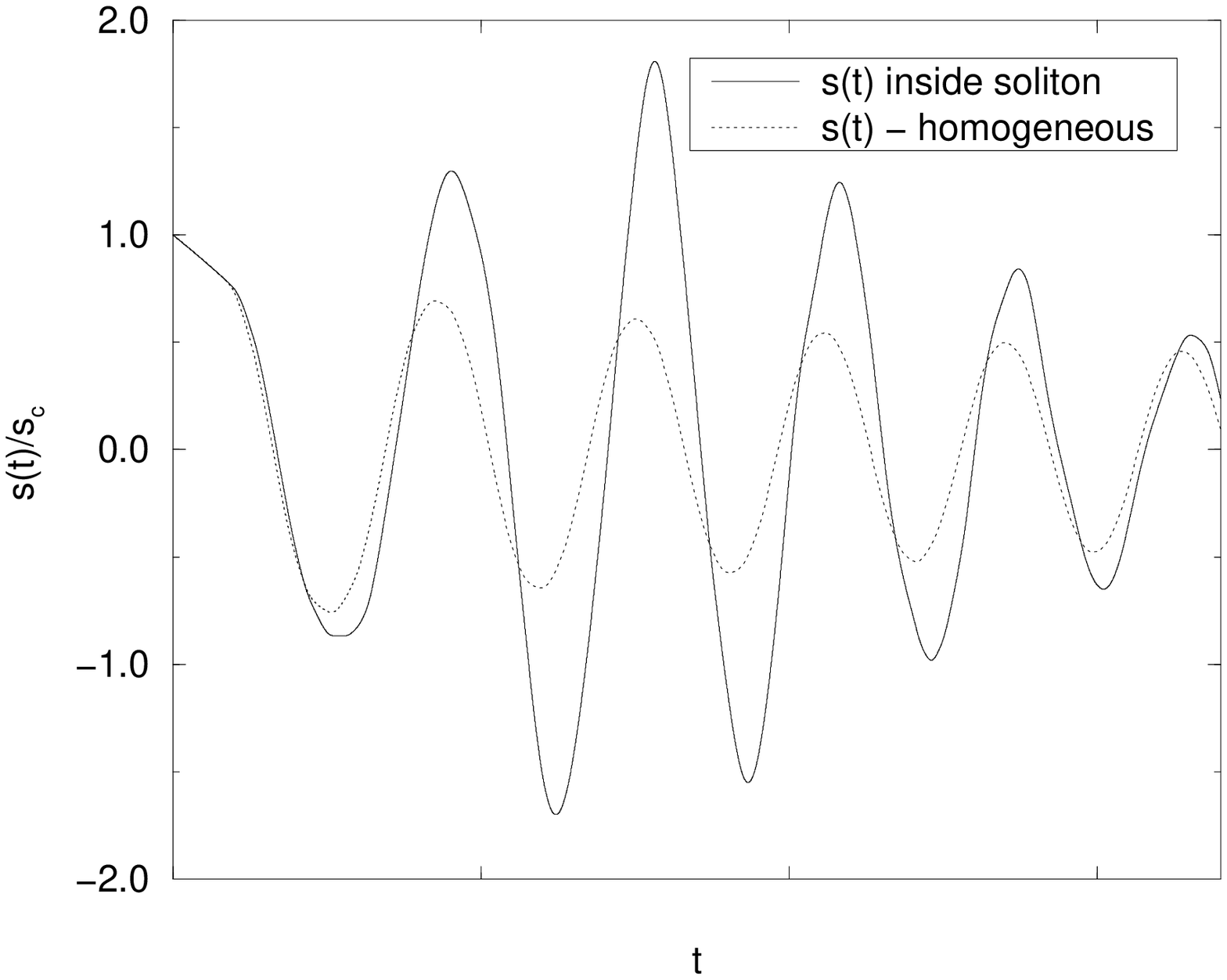}
\caption{\footnotesize{Coherent oscillations of an energy density lump.}}
\label{fig:fig1}
\end{center}
\end{figure}

              In Figure 3 we show the results of a 3-D simulation on a $50^{3}$
lattice. The regions where the energy density is greater than $4 \overline{\rho}$ are shown. 
This clearly demonstrates the formation of inflaton condensate lumps in the 3-D case,
confirming the results of the higher resolution 2-D simulations.  

\begin{figure}[htbp]
\begin{center}
\includegraphics[width=0.75\textwidth]{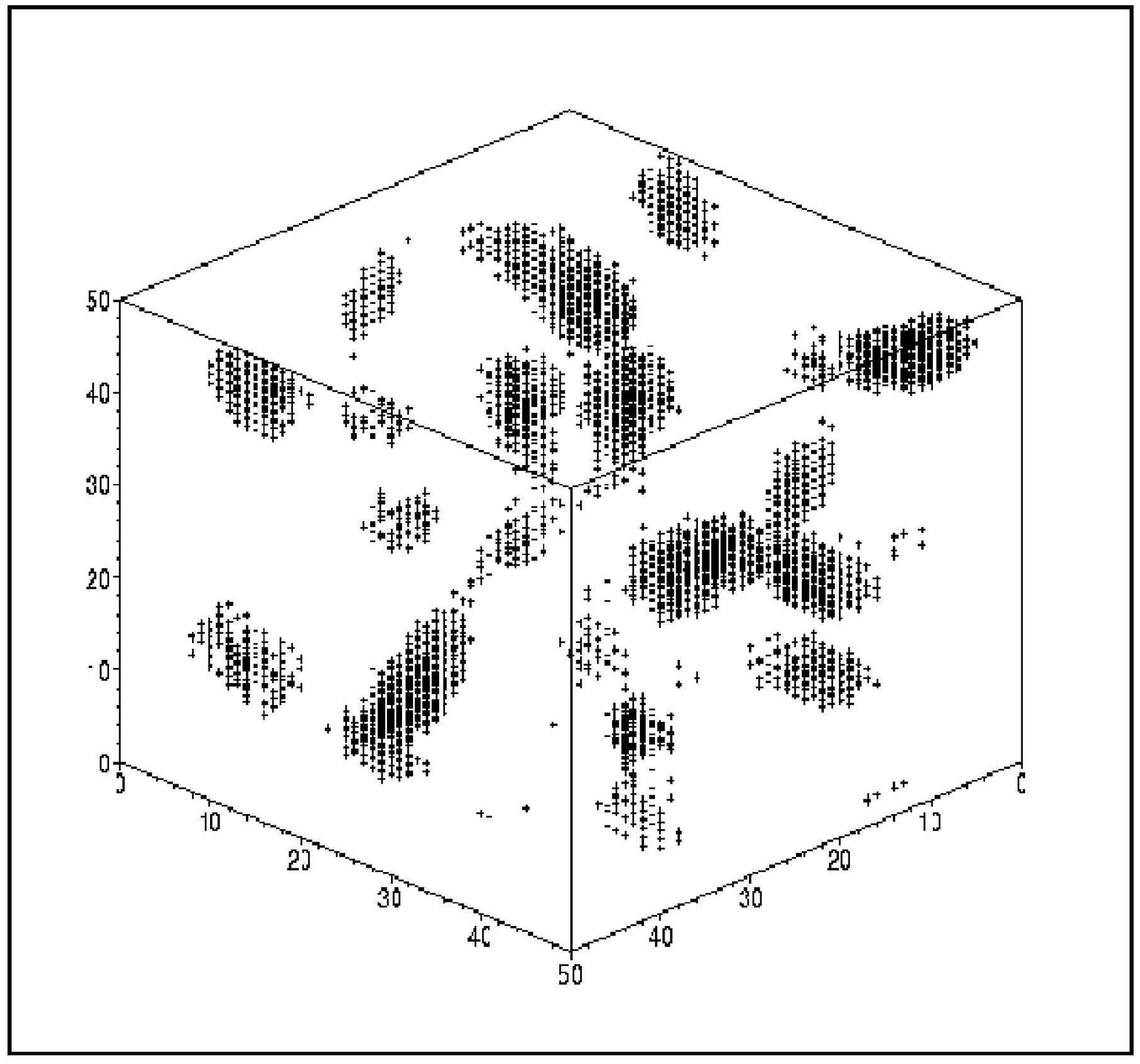}
\caption{\footnotesize{ $50^{3}$ lattice 3D simulation. 
The plot shows regions of energy $\rho > 4 \overline{\rho}$
at $t = 5 \tau_{S}$. }}
\label{fig:fig1}
\end{center}
\end{figure}

    We find no evidence of cosmic strings forming within the volume
of our our lattice. This is consistent with our previous semi-analytical 
analysis, which indicated that the typical seperation of $U(1)_{FI}$ cosmic strings formed 
at the end of hybrid inflation would be not very much smaller than the horizon, corresponding to the 
wavelength of the first $\Phi_{-}$ fluctuations to be able to 
grow classically and break the $U(1)_{FI}$ symmetry \cite{mb1}.

\section{A Scaling Property of the D-term Inflation Field Equations.}

    Throughout we have considered the case $g = 1$. However,  
by rescaling the spacetime coordinates, 
the D-term inflation field equations can be written in a form 
that depends only upon the ratio $\lambda/g$. As a result, 
the condition for the formation of inflaton condensate lumps is likely to depend 
primarily upon the ratio $\lambda/g$, with only a 
weak logarithmic dependence on $g$ due to the initial value of the 
perturbations. 

       Under the rescaling of the coordinates
$t = \tilde{t}/g$, $x = \tilde{x}/g$, the tree-level scalar field equations become 
\be{at1} \ddot{S} + 3 \tilde{H} \dot{S} - \frac{\vec{\nabla}^{2}}{a^{2}}S    
= -\frac{\lambda^{2}}{g^{2}}|\Phi_{-}|^{2}S      ~,\ee
and
\be{at2} \ddot{\Phi}_{-} + 3 \tilde{H} \dot{\Phi}_{-} - \frac{\vec{\nabla}^{2}}{a^{2}}\Phi_{-}    
= -\frac{\lambda^{2}}{g^{2}} |S|^{2}\Phi_{-} + \left(\xi - |\Phi_{-}|^{2} \right)
\Phi_{-}     ~,\ee
where derivatives are now with respect to $\tilde{x}$ and $\tilde{t}$. We have defined 
$\tilde{H} = a^{-1}da/d\tilde{t} = H/g$. $\tilde{H}$ is a function purely of 
$\lambda/g$. This follows since $H^{2} \propto \rho$ where, in terms of derivatives with respect to 
rescaled spacetime coordinates,
\be{at3} \rho = g^{2}\left(|\dot{S}|^{2} + |\dot{\Phi}_{-}|^{2} + \left|\frac{\nabla S}{a}\right|^{2} 
+ \left|\frac{\nabla \Phi_{-}}{a}\right|^{2} + V(S, \Phi_{-})/g^{2}   \right)  \equiv g^{2} \tilde{\rho}  ~.\ee 
$V(S, \Phi_{-})/g^{2}$ is purely a function of $\lambda/g$, therefore $\tilde{\rho}$ depends only on 
$\lambda/g$. Thus $H \propto \rho^{1/2} = g \tilde{\rho}^{1/2}$ and so $\tilde{H} = H/g \propto \tilde{\rho}^{1/2}$. 
Therefore $\tilde{H} = \tilde{H}(\lambda/g)$.

    The dependence of the rescaled field equations only on $\lambda/g$ does not extend to the 
quantum fluctuation initial conditions, \eq{fe7}. In terms of rescaled coordiantes and 
the rescaled wavenumber, $\tilde{k} = k/g$, the initial fluctuation becomes 
\be{at3a}  \delta \phi(\tilde{\vec{x}}, 0)  = 
\frac{1}{\sqrt{V}} \sum \frac{8g^{-1/2}}{\sqrt{2 \tilde{k}_{\alpha \beta \gamma}}} 
\cos\left( \Delta \tilde{k} \Delta \tilde{x} \alpha i + \theta_{\alpha} \right) 
\cos\left( \Delta \tilde{k} \Delta \tilde{x} \beta j + \theta_{\beta} \right)
\cos\left( \Delta \tilde{k} \Delta \tilde{x} \gamma k + \theta_{\gamma} \right)   ~.\ee
The upper limit of the sum, corresponding to the largest value of $\tilde{k}$ for which the modes can grow,  
is determined from the rescaled field equations and is therefore purely a function of $\lambda/g$. Thus the 
initial conditions for the rescaled field equations are proportional to $g^{-1/2}$.  
However, the growth of perturbations in a tachyonic potential 
is typically rapid (exponential), in which case the condition for
perturbations to become non-linear will depend only weakly (logarithmically) on the initial perturbation. 
As a result, if inflaton condensate lumps form for 
one value of $g$ (we considered $g = 1$ in our simulations) then they will form for any value of $g$  
for the same value of $\lambda/g$, up to a weak dependence on the initial conditions.  
The effect of reducing $g$ will be to increase the time for the lumps to form 
($t = \tilde{t}/g$) and to increase the size of the resulting lumps ($x = \tilde{x}/g$).

\section{Comparison with Previous Simulations}

    In \cite{tp1} a simulation of SUSY hybrid inflation was reported for the case equivalent to
D-term inflation couplings $\lambda = 0.014$ and $g = 0.14$, such that $\lambda/g = 0.1$. 
It was concluded that the 
energy density would be transferred to inhomogeneous classical scalar field waves. Although no 
non-topological soliton formation was observed, this value of $\lambda/g$ is on the borderline of values
for which we would expect to find non-topological soliton formation \cite{mb1}. 
Therefore the results of \cite{tp1} are not in any obvious way inconsistent with the results we have obtained here.   

In \cite{cpr} a simulation of the end of D-term 
was performed, again for the case equivalent to $\lambda = 0.014$, $g=0.14$ in D-term inflation. 
The formation of non-topological solitons
was observed along the length of cosmic strings. 
However, only quantum fluctuations of the symmetry-breaking field were 
considered in \cite{cpr}, with no inflaton fluctuations, which cannot be considered fully realistic. 
As discussed above, we expect that fluctuations of the symmetry-breaking field will only play a 
minor role in the evolution of the total energy density, being primarily associated with 
cosmic-string formation on scales not very much smaller than the horizon \cite{mb1}. 
Nevertheless, the results in \cite{cpr} confirm the possibility of non-topological soliton
formation at the end of SUSY hybrid inflation.

\section{Conclusions and Discussion}

      Our numerical simulations have shown that growth of quantum fluctuations
of the scalar fields in SUSY hybrid inflation models results in the
formation of inflaton condensate lumps corresponding to 
spherical lumps of coherently oscillating scalar field, also known as oscillons. 
The energy density of the Universe becomes almost
entirely concentrated in these condensate lumps. 
We have focused on the case of D-term hybrid inflation,
in which case the inflaton condensate lumps form if $\lambda \gae 0.1 g$. 
Although a full analysis remain to be done, similar
results may be expected in the case of F-term inflation, whose classical 
field equations are similar to those of D-term inflation for the case $\lambda = \sqrt{2} g$.

        We have performed simulations for the case of a real inflaton field. 
The fact that the objects which form in this case are oscillons is significant,
since from numerical and analytical studies of single-field oscillons such objects are known
to be quasi-stable, with very long lifetimes compared to the scalar field coherent oscillation period.
Therefore we expect that the post-inflation era will consist entirely of long-lived
inflaton condensate lumps, even though our present numerical simulations are unable to resolve
the evolution over more than about 10 coherent oscillation periods. 

        In the case of a complex inflaton field, oscillons are expected
to be a precursor to stable Q-ball formation via decay to Q-ball pairs.
Decay to Q-ball pairs is likely to occur long before the  
neutral inflaton condensate lumps decay via radiation of scalar field waves.
Thus if inflaton condensate lumps form in the case of a
real inflaton field, it is very likely that decay to Q-ball pairs and 
a Q-ball dominated post-inflation era will occur once a complex inflaton is considered. 
Since the Q-balls are classically stable, this era will only end once the 
scalar particles forming the Q-ball decay perturbatively to MSSM
particles, reheating the Universe. A highly inhomogeneous inflaton condensate lump or Q-ball
dominated era would have significant consequences for post-inflation SUSY cosmology.
For example, it has been shown that the dynamics of scalar fields in the post-inflation era
will not be subject to order $H$ mass corrections \cite{fdd}, with important consequences for
Affleck-Dine baryogenesis, SUSY curvatons and moduli dynamics. Reheating
will also be inhomogeneous, via Q-ball decay.
          
     We plan to develop  
the capabilities of our numerical simulations in the future, 
with the ultimate goal of achieving a complete understanding of the post-inflation era of 
SUSY hybrid inflation models.

\section*{Acknowledgements}   The work of MB was supported by EPSRC.

\renewcommand{\theequation}{A-\arabic{equation}}
 \setcounter{equation}{0} 

\section*{Appendix.  Non-linear Growth of Quantum Fluctuations
 and the Equivalent Classical Stochastic Field}

            In this Appendix we review the growth of quantum fluctuations in
the presence of a time-dependent tachyonic mass squared term. 
We first consider a real scalar field in Minkowski space with 
a positive or negative mass squared term. We
then generalise to the case of a Friedmann-Robertson-Walker
background, in particular de Sitter space. 

               In order to derive the classical initial conditions used
in the lattice simulations, we quantize the theory in a box of side $L$. 
We consider an initial state corresponding to a Minkowski or Bunch-Davies vacuum state
for a massive scalar field. The scalar field theory is exactly quantized in terms of
creation and annihilation operators for momentum modes, which is valid so long as interaction
terms can be neglected in the Hamiltonian, or equivalently the scalar field equations are linear.
To follow the evolution of the quantum fluctuations into the non-linear regime, the
quantum field theory is replaced by an equivalent classical field theory with a classical probability
distribution for the initial conditions \cite{guth,ps}, equivalent in the sense that the
results of quantum field theory for the expectation values of products of fields are equal to the values
obtained by solving the classical field equations with the classical probability distribution
for the initial field values.  This can be done if the occupation numbers of
the quantum modes become large compared with one during the linear regime. 
The classical probability distribution is then given by the Wigner
function of the field modes in the semi-classical limit. Although the classical
probability distribution must become
valid at some time during the period when the theory can be considered linear,
once it becomes established it can be applied at any earlier time during the linear regime.

        The Lagrangian and Hamiltonian for the scalar field in Minkowski space are
\be{a0} {\cal L} = \frac{1}{2}\partial_{\mu}\phi \partial^{\mu} \phi  - \frac{m^{2}}{2} \phi^{2}    ~,\ee
and
\be{a1} {\cal H} = \pi \dot{\phi} - {\cal L}
 = \frac{1}{2} \left[ \pi^{2} + ( \vec{\nabla} \phi)^{2} + m^{2} \phi^{2} \right]     ~,\ee
where the canonical momentum is $ \pi 
\equiv \partial {\cal L}/ \partial \dot{\phi} = \dot{\phi}$. We consider
the scalar field in a cube of volume $L^3$ with periodic boundary conditions.
The field and canonical momentum are expanded in terms of momentum modes as 
\be{a2}  \phi(\vec{x}, t) = \frac{1}{\sqrt{V}} \sum_{\vec{k}} q(\vec{k}, t) e^{i \vec{k}. \vec{x}}    ~\ee
and
\be{a3}  \pi(\vec{x},t) = \frac{1}{\sqrt{V}} \sum_{\vec{k}} p(\vec{k}, t) e^{i \vec{k}.\vec{x}}   ~,\ee
where the components of $\vec{k}$ satisfy $k_{i} = 2 \pi n_{i}/L$ with $n_{i}$ an integer. 
$\pi(\vec{x},t) = \dot{\phi}(\vec{x},t)$ implies that $p(\vec{k}, t)
 = \dot{q}\left(\vec{k},t\right) $.
The non-zero equal-time canonical commutation relations for the field operators are 
\be{a3a}  [\pi(\vec{x},t), \phi(\vec{x}^{'},t)] = -i \delta^{3}\left(\vec{x} - \vec{x}^{'}\right) 
~.\ee    
The field and conjugate momentum mode operators must then satisfy 
\be{a4} \left[p(-\vec{k},t), q(\vec{l},t)\right] = -i \delta_{\vec{k},\vec{l}}    ~.\ee  
In terms of mode operators the Hamiltonian becomes  
\be{a5}  H = \frac{1}{2} \sum_{\vec{k}} p^{\dagger}(\vec{k},t) p(\vec{k},t) + \omega_{\vec{k}}^{2} 
q^{\dagger}(\vec{k},t) q(\vec{k},t) \;\;,\;\;\; \omega_{\vec{k}}^{2} = \vec{k}^{2} + m^{2}  
~,\ee
where the mode operators satisfy the classical Hamiltonian equations of motion.   
In general, a Hamiltonian which is quadratic in the mode operators may be exactly quantized by expanding the 
 operators in terms of time-independent creation and annihilation operators 
\be{a6}    q(\vec{k},t) = f_{\vec{k}} a_{\vec{k}}
 + f_{\vec{k}}^{\dagger} a_{-\vec{k}}^{\dagger}      ~\ee
and 
\be{a6a}    p(\vec{k},t) = - i g_{\vec{k}} a_{\vec{k}}
 + i g_{\vec{k}}^{\dagger} a_{-\vec{k}}^{\dagger}      ~\ee
where $[a_{\vec{k}}, a_{\vec{l}}^{\dagger}] = \delta_{\vec{k},\vec{l}}  $. 
$p(\vec{k},t) = \dot{q}(\vec{k},t)$ implies that $g_{\vec{k}} = i \dot{f}_{\vec{k}}$. 
The mode functions $f_{\vec{k}}$ and $g_{\vec{k}}$ satisfy the classical field equations for the
 momentum modes (which are, by assumption, linear in $f_{\vec{k}}$) and the canonical commutation relations.
 The latter implies that the mode functions satisfy the condition
 $g_{\vec{k}} f_{\vec{k}}^{\dagger} + g_{\vec{k}}^{\dagger} f_{\vec{k}} = 1$. 
 For the case of a constant positive mass squared term, the mode functions $f_{\vec{k}}$ must
 also satisfy the condition that the 
Hamiltonian is correctly quantized, such that the energy eigenstates correspond to numbers
 of quanta of energy $\omega_{\vec{k}}$. 
The unique solution for $f_{\vec{k}}$ which satisfies both of these conditions is 
\be{a7}   f_{\vec{k}}  = \frac{i}{ \sqrt{2 \omega_{\vec{k}} }} e^{-i \omega_{\vec{k}}t}  ~.\ee
 To study the evolution of the mode functions from an initial time $t=0$ at which $m^2 \geq 0$,
 we use \eq{a7} as the initial condition for the mode functions. The subsequent evolution
 may then be studied by solving the classical equation of motion for $f_{\vec{k}}$ so long as we are
 in the linear regime, since in this case the 
 field equations may be satisfied by field operators expanded in terms of time-independent
 creation and annihilation operators.

     As an example, we consider the evolution for a model where a field
is massless at $t \leq 0$ and gains a constant negative mass squared term once $t > 0$. 
The field operator at $t > 0$ satisfies the equation, for modes with $k^2 < |m|^2$, 
\be{a8}  \ddot{q}\left(\vec{k},t\right) = \tilde{\omega}_{\vec{k}}^{2} q\left(\vec{k},t\right) 
 \;,\;\;\; \tilde{\omega}^{2}_{\vec{k}} = |m|^{2} - \vec{k}^{2}    ~,\ee
with general solution
\be{a9}  q(\vec{k},t) = \alpha_{\vec{k}} e^{\tilde{\omega}_{\vec{k}} t}
 + \beta_{\vec{k}} e^{-\tilde{\omega}_{\vec{k}} t}        ~\ee
and
\be{a9a}  p(\vec{k},t)\equiv \dot{q}(\vec{k},t) = \tilde{\omega}_{\vec{k}}\left( 
\alpha_{\vec{k}} e^{\tilde{\omega}_{\vec{k}} t}
 - \beta_{\vec{k}}  e^{-\tilde{\omega}_{\vec{k}} t} \right)        ~,\ee
where $\alpha_{\vec{k}}$ and $\beta_{\vec{k}}$ are time-independent operators corresponding
to linear combinations of $a_{\vec{k}}$ and $a_{-\vec{k}}^{\dagger}$. Matching these with the 
vacuum ($m = 0$) solution at $t = 0$ implies that
\be{a10} \alpha_{\vec{k}} =
\frac{i}{2 \sqrt{2 \omega_{\vec{k}}  }} \left[ \left(1 - \frac{ i\omega_{\vec{k}} }{\tilde{\omega}_{\vec{k}}}
\right) a_{\vec{k}} - 
\left(1 +  \frac{i \omega_{\vec{k}}}{\tilde{\omega}_{\vec{k}}}
\right) a_{-\vec{k}}^{\dagger} \right]       ~\ee
and 
\be{a11} \beta_{\vec{k}} =
 \frac{i}{2 \sqrt{2 \omega_{\vec{k}}  } }\left[ \left(1 +  \frac{i\omega_{\vec{k}}}{\tilde{\omega}_{\vec{k}}}
\right) a_{\vec{k}}
 - \left(1 - \frac{i \omega_{\vec{k}}}{\tilde{\omega}_{\vec{k}}}
\right) a_{-\vec{k}}^{\dagger} \right]       ~,\ee
where $\omega_{\vec{k}} = |\vec{k}|$. 
Therefore $q(\vec{k},t)$ for $t > 0$ can be written in the form of \eq{a6} with 
\be{a11a} f_{\vec{k}} =  \frac{i}{2 \sqrt{2 \omega_{\vec{k}} 
 } }\left[ \left(1 - \frac{i \omega_{\vec{k}}}{\tilde{\omega}_{\vec{k}}}
\right) e^{\tilde{\omega}_{\vec{k}} t}
 + \left(1 + \frac{i \omega_{\vec{k}}}{\tilde{\omega}_{\vec{k}}}
\right)e^{-\tilde{\omega}_{\vec{k}} t}  \right]    ~.\ee
For $\tilde{\omega}_{\vec{k}} t \gg 1$, 
$q(\vec{k},t) \approx \alpha_{\vec{k}}e^{\tilde{\omega}_{\vec{k}} t}$ and 
$p(\vec{k},t) \approx \omega_{\vec{k}} \alpha_{\vec{k}}e^{\tilde{\omega}_{\vec{k}} t}$, 
so that to a good approximation $q(\vec{k},t)$ and $p(\vec{k},t)$ commute, corresponding to the limit
of classical physics.

    In order to study the growth of quantum fluctuations beyond the linear regime, we need
the classical stochastic initial conditions which reproduce the 
field theory expectation values in the semi-classical limit. The classical probability
distribution is obtained from the Wigner function, 
\be{a15}    W(q,p,t) = \int \frac{dx_{1} dx_{2}}{\left( 2 \pi \right)^{2}} 
e^{ -i \left( p_{1} x_{1} + p_{2} x_{2} \right)} < q-\frac{x}{2}, t| \rho |q + \frac{x}{2}, t> 
~,\ee 
where $p =p_{1} + i p_{2}$, $q =q_{1} + i q_{2}$ and $x =x_{1} + i x_{2}$. The $q$ and $p$ represent
complex generalised coordinates and conjugate momenta, which in our case correspond to 
$q(\vec{k},t)$ and $p(\vec{k},t)$. $\rho = |\psi_{o}(t)><\psi_{o}(t)|$
is the density matrix of the Schrodinger picture state, 
$|\psi_{o}(t)>$, which evolves from the initial vacuum state. 
In general the Wigner function gives 
the probability distribution of the observable $q$ at $t$
after integrating over $p$ and vice-versa. 
In the classical limit the Wigner function tends towards a classical probability distribution, 
$f(q,p,t)$, which reproduces the quantum field theory results for expectation values of the
field operators. Although $f(q,p,t)$ should be calculated
in the semi-classical limit, it can be applied at any time during the linear evolution
of the field, in particular at earlier times, since it satisfies the
classical equations of motion \cite{guth}. 
In the case of hybrid inflation we can therefore apply the classical
initial conditions at the end of inflation ($s = s_{c}$), 
corresponding to massless scalar fields.

      The annihilation operators can be expressed as 
\be{a14} a_{\vec{k}} = \frac{g_{\vec{k}}^{\dagger}}{2 Re(f_{\vec{k}} g_{\vec{k}}^{\dagger})}
\left(q(\vec{k},t) - \frac{f_{\vec{k}}^{\dagger}}{i g_{\vec{k}}^{\dagger}} p(\vec{k},t) 
\right)      ~.\ee
By definition $a_{\vec{k}} |0> = 0$,  where the Heisenberg picture initial vacuum state, $|0>$,
is equivalent to the Schrodinger picture vacuum state at $t=0$. 
In order to obtain the Schrodinger wavefunction $\psi(q,t) \equiv <q|\psi_{o}(t)>$ corresponding 
to the state evolving from the initial vacuum state, we express the equation $a_{\vec{k}} |0>$
as a wave equation in the position representation \cite{ps}. In terms of the time evolution 
operator $U(t)$, $a_{\vec{k}} |0> = 0$ can be expressed as 
\be{a18} U^{\dagger} [q(\vec{k},0) + \frac{i f_{k}^{\dagger}}{g_{k}^{\dagger}} p(\vec{k},0) ] U |0>  = 0      ~,\ee
where $q(\vec{k},t) = U^{\dagger} q(\vec{k},0) U$. Since $U|0>  \equiv |\psi_{o}(t)>$ is the time-dependent
Schrodinger state vector and $q(\vec{k},0) \equiv q_{\vec{k}}$ , $p(\vec{k},0) \equiv p_{\vec{k}}$ are the 
Schrodinger representation generalised coordinate and momentum operators, with $p_{\vec{k}}$ 
 conjugate to $q_{-\vec{k}} \equiv q^{\dagger}_{\vec{k}}$ ($[p_{\vec{k}}, q_{\vec{k}}^{\dagger}] = -i$), 
the momentum operator in the position representation becomes 
$p_{\vec{k}} = -i \partial / \partial q_{\vec{k}}^{*}$. The wave equation can then be written as 
\be{a17} \left(q_{\vec{k}} 
+ \frac{f_{\vec{k}}^{\dagger}}{g_{\vec{k}}^{\dagger}} \frac{\partial}{\partial q_{\vec{k}}^{*}} \right) 
\psi_{o}\left(q_{\vec{k}}, q_{\vec{k}}^{*}, t \right)  = 0    ~.\ee 
This has the solution \cite{ps}
\be{a18a}  \psi_{o}(q_{\vec{k}}, q_{\vec{k}}^{*}, t) = N_{o} \exp\left( - \frac{g_{k}^{\dagger}}{f_{k}^{\dagger}} 
|q_{\vec{k}}|^{2}\right) ~,\ee
where $N_{o} = (\sqrt{\pi} |f_{\vec{k}}|)^{-1}$. The normalized probability distribution is then 
\be{a19} P(q_{\vec{k}}, q_{\vec{k}}^{*}, t) = \frac{1}{ \pi |f_{\vec{k}}|^{2}}
 exp\left( - \frac{|q_{\vec{k}}|^{2}}{|f_{\vec{k}}(t)|^{2}} \right)     ~.\ee
This gives the probability that the magnitude of the {\it classical} mode amplitude 
$q(\vec{k}, t)$ at $t$ is $|q_{\vec{k}}|$. This is a complex Gaussian distribution, 
with r.m.s. value for $|q_{\vec{k}}|$ equal to $|f_{\vec{k}}|$. Then 
$q(\vec{k}, t) = |q(\vec{k}, t)| e^{i \delta_{\vec{k}}}$, with $\delta_{\vec{k}}$ 
a random phase.

                     The Wigner function is then given by    
\be{a20} W(q_{\vec{k}}, p_{\vec{k}}, t) = \int \int \frac{ dx_{1} dx_{2}}{\left( 2 \pi\right)^{2}} 
e^{-i \left(p_{\vec{k} \;1} x_{1} + p_{\vec{k}\;2} x_{2}\right)} \psi_{o}^{*}(q_{\vec{k}}-x/2, t)
 \psi_{o}(q_{\vec{k}} + x/2, t) 
~.\ee
Integrating gives 
  \be{a21} W(q_{\vec{k}},p_{\vec{k}},t) = \frac{1}{\pi^{2}} \exp\left[ -\frac{|q_{\vec{k}}|^{2}}{|f_{\vec{k}}|^{2}} - 
|f_{\vec{k}}|^{2} \left|p_{\vec{k}} - \frac{F_{\vec{k}}}{|f_{\vec{k}}|^{2}} q_{\vec{k}}\right|^2 \right]          ~,\ee
where $F_{\vec{k}} = Im(f_{\vec{k}}^{\dagger}g_{\vec{k}})$.  In the limit  $|f_{\vec{k}}|^{2} \gg 1$, where the Wigner
 function tends towards the classical probability distribution, the Wigner function is
 vanishing except if 
\be{a22} p_{\vec{k}} = \frac{F_{\vec{k}}}{|f_{\vec{k}}|^{2}} q_{\vec{k}}   ~.\ee
Since the classical probability distribution can be applied at any time during the linear regime,
we can choose to apply the initial conditions at the end of hybrid inflation, such that
the fields are massless. For the case of flat space, 
the mode functions with $m = 0$ are given by
\be{a23}   f_{\vec{k}} = \frac{i}{\sqrt{2k}} e^{-i kt} 
\;,\;\;\; g_{\vec{k}} = i \sqrt{\frac{k}{2}} e^{-i kt}
   ~,\ee
where $k = |\vec{k}|$. 
Thus $F_{\vec{k}} = Im(f_{\vec{k}}^{\dagger}g_{\vec{k}}) = 0$. 
Therefore the flat space classical initial condition for $p(\vec{k},t)$
is $p(\vec{k},0) = 0$, whilst the
flat space classical initial conditions for $q(\vec{k},t)$
at $t=0$ corresponds to  
a complex Gaussian distribution for $|q(\vec{k},0)|$,
with r.m.s. value   
\be{a24} |q(\vec{k},0)|_{rms} =  \frac{1}{\sqrt{2 k}}   ~\ee
and with a random phase factor $e^{i \delta_{\vec{k}}}$ for each $\vec{k}$.

       These results can easily be generalised to a scalar field in  
Friedman-Robertson-Walker spacetime. In this case the action is  
\be{a23a} S = \int d^{4} x \sqrt{-g} \left[\frac{1}{2} \partial_{\mu}\phi \partial^{\mu}\phi 
 - \frac{m^{2}}{2} \phi^{2} \right]    ~.\ee
In terms of $y = a(t) \phi$ and conformal time $\eta$ ($dt = a(t)d \eta$), where $a(t)$ is 
the scale factor, this becomes
\be{a24a} S = \int d^{4} x \left[ \frac{1}{2} \left(y^{'} - \frac{a^{'}}{a} y \right)^{2} 
- \frac{\left(\nabla y\right)^{2}}{2} - \frac{m^2 a^2}{2} y^2      \right]      ~,\ee
where $'$ denotes differentiation with respect to conformal time and $d^{4} x$ is now understood 
as an integral over comoving coordinates and conformal time. 
The conjugate momentum is 
\be{a25}   \pi = \frac{\partial {\cal L}}{\partial y^{'}} = y^{'} - \frac{a^{'}}{a}y  ~\ee
and the Hamiltonian is  
\be{a26} H = \int d^{3}x \left[ \frac{\pi^{2}}{2} + \frac{\left(\nabla y\right)^{2}}{2}
 + \frac{m^2 a^2}{2} y^2   + \frac{a^{'}}{a}\pi y \right]  ~.\ee
Expanding $y(\vec{x},\eta)$ and $\pi(\vec{x},\eta)$ in terms of 
modes as before (but now with $\eta$ in place of $t$ and $y$ in place of $\phi$) 
implies that
\be{a27} H = \frac{1}{2} \sum_{\vec{k}} \left[p(\vec{k},\eta)^{\dagger}p(\vec{k},\eta) 
+ \omega_{\vec{k}}^{2}  q(\vec{k},\eta)^{\dagger}q(\vec{k},\eta) + \frac{a^{'}}{a} 
\left(p(\vec{k},\eta)q^{\dagger}(\vec{k},\eta) + p^{\dagger}(\vec{k},\eta)q(\vec{k},\eta) \right) \right] 
~,\ee
where $\omega_{\vec{k}}^{2} = \vec{k}^2 + m^2 a^2$ and
\be{a28a} p(\vec{k},\eta) = q^{'}(\vec{k},\eta) - \frac{a^{'}}{a}q(\vec{k},\eta)    ~.\ee  
This is quantized by expanding $q$ and $p$ in terms of creation and annihilation operators 
according to \eq{a6} and \eq{a6a}. 
The mode functions satisfy the classical equation of motion 
\be{a29} f_{\vec{k}}^{''} - \frac{a^{''}}{a} f_{\vec{k}} = -\left( \vec{k}^{2} +m^2 a^2\right) f_{\vec{k}}    ~,\ee
with the requirement that the initial $f_{\vec{k}}$ reduces to the Minkowski modes in the limit  
$k \gg H$ (Bunch-Davies vacuum).    

         For the case of de Sitter space, $H = constant$, the mode equation becomes 
\be{a29a} f_{\vec{k}}^{''}  + \left[k^{2} - \frac{ \left(2 - \frac{m^{2}}{H^{2}} \right)}{
\eta^{2}  } \right]  f_{\vec{k}} = 0     ~,\ee
where $\eta = -1/aH$.  
For a constant $m^2$, the solution which tends to the Minkowski vacuum in the limit $k \gg H$ is
\be{a30}    f_{\vec{k}}(\eta) = \frac{\sqrt{\pi}}{2} e^{i \left( \frac{\pi}{2}\nu + \frac{\pi}{4}\right)}
\sqrt{-\eta} H_{\nu}^{(1)}\left(-k\eta\right) \;,\;\;\; \nu^{2} = \frac{9}{4} - \frac{m^{2}}{H^{2}}      ~,\ee
where $H^{(1)}_{\nu}$ is a spherical Hankel function of the first kind. 

       The Wigner function analysis is as before except with $\eta$
in place of $t$. $f_{\vec{k}}$ is now given by \eq{a30}, whilst 
$g_{\vec{k}}$ follows from $p(\vec{k},\eta) = q^{'}(\vec{k},\eta) - (a^{'}/a) q(\vec{k},\eta)$, which 
implies that $g_{\vec{k}} = i (f_{\vec{k}}^{'} - (a^{'}/a)
 f_{\vec{k}})$.  The initial 
conditions we impose correspond to de Sitter space with a massless field, in which case $\nu = 3/2$ and so
\be{a31} f_{\vec{k}} = \frac{i \sqrt{\pi}}{2} \sqrt{-\eta} H_{\frac{3}{2}}^{(1)} \left( -k \eta\right) 
 = \frac{e^{-i k \eta}}{\sqrt{2k}} \left[1 - \frac{i}{k \eta} \right]      ~\ee 
\be{a32} g_{\vec{k}} = \sqrt{\frac{k}{2}}  e^{-i k\eta}        ~,\ee
where we have used the expression for the Hankel function, 
\be{a32a} H_{3/2}^{\left(1\right)}(x) = -\left(\frac{2}{\pi x}\right)^{1/2} e^{i x}
\left[1 + \frac{i}{x} \right]     ~.\ee
Thus $F_{\vec{k}} = Im(f_{\vec{k}}^{\dagger}g_{\vec{k}}) = 1/(2 k \eta)$. 
The classical r.m.s. values for the magnitudes of the mode functions are then 
\be{a34} |q(\vec{k},\eta)|_{rms}  = |f_{\vec{k}}| = 
\frac{1}{\sqrt{2 k}} \left( 1 + \frac{1}{k^{2} \eta^{2}} \right)^{1/2}    ~\ee
and
\be{a35} |p(\vec{k},\eta)|_{rms} = \frac{1}{\sqrt{2 k} |\eta|} \left(1 
+ \frac{1}{k^{2} \eta^{2}} \right)^{-1/2}     ~.\ee
These may be used to model the initial conditions for the massless scalar fields at the
end of hybrid inflation. 

           In practice, we numerically solve the classical 
scalar field equations for $\phi(\vec{x},t)$ rather than $y(\vec{x},\eta)$.
In general, $y(\vec{x},\eta) = a \phi(\vec{x},t)$ and 
$\pi(\vec{x},\eta) = a^{2} \dot{\phi}(\vec{x},t)$. 
Choosing $a = 1$ at the end of inflation (corresponding to choosing $\eta = -H^{-1}$ at $t=0$)  
implies that $\phi(\vec{x},0) = y(\vec{x},\eta)$ and $\dot{\phi}(\vec{x},0) = 
\pi(\vec{x},\eta)$ at the end of inflation.  In this case the
mode functions in \eq{a34} and \eq{a35} will provide the initial conditions for the 
mode expansions of $\phi(\vec{x},t)$ and $\dot{\phi}(\vec{x},t)$ in de Sitter space.

\end{document}